\documentclass[3p,twocolumn]{elsarticle}
\usepackage{graphicx}
\usepackage{epsfig}
\usepackage{amsmath}
\usepackage{dcolumn}
\usepackage{bm}
\usepackage{dcolumn}
\usepackage{color}
\usepackage{verbatim}
\usepackage{paralist}
\usepackage{amssymb}
\usepackage{url}
\begin{document}
\newcommand{\exposure}{252.6\,ton$\cdot$yr}
\newcommand{\unit}{events/(100\,ton$\cdot$yr)}
\begin{frontmatter}
\title{Observation of Geo--Neutrinos} 
\author[Milano]{G.~Bellini}
\author[PrincetonChemEng]{J.~Benziger}
\author[Milano]{S.~Bonetti}
\author[Milano]{M.~Buizza~Avanzini}
\author[Milano]{B.~Caccianiga}
\author[UMass]{L.~Cadonati}
\author[Princeton]{F.~Calaprice}
\author[Genova]{C.~Carraro}
\author[Princeton]{A.~Chavarria}
\author[Princeton]{F.~Dalnoki-Veress\fnref{label1}}
\fntext[label1]{At present at James Martin Center for Nonproliferation Studies
Monterey Institute of International Studies.}
\author[Milano]{D.~D'Angelo}
\author[Genova]{S.~Davini}
\author[APC]{H.~de~Kerret}
\author[Peters]{A.~Derbin}
\author[Kurchatov]{A.~Etenko}
\author[Ferrara]{G.~Fiorentini}
\author[Dubna]{K.~Fomenko}
\author[Milano]{D.~Franco}
\author[Princeton]{C.~Galbiati\fnref{label2}}
\fntext[label2]{Also at Fermi National Accelerator Laboratory.}
\author[LNGS]{S.~Gazzana}
\author[LNGS]{C.~Ghiano}
\author[Milano]{M.~Giammarchi}
\author[Munich]{M.~Goeger-Neff}
\author[Princeton]{A.~Goretti}
\author[Genova]{E.~Guardincerri}
\author[Virginia]{S.~Hardy}
\author[LNGS]{Aldo~Ianni}
\author[Princeton]{Andrea~Ianni}
\author[Virginia]{M.~Joyce}
\author[Kiev]{V.V.~Kobychev}
\author[LNGS]{Y.~Koshio}
\author[LNGS]{G.~Korga}
\author[APC]{D.~Kryn}
\author[LNGS]{M.~Laubenstein}
\author[Princeton]{M.~Leung}
\author[Munich]{T.~Lewke}
\author[Kurchatov]{E.~Litvinovich}
\author[Princeton]{B.~Loer}
\author[Milano]{P.~Lombardi}
\author[Milano]{L.~Ludhova}
\author[Kurchatov]{I.~Machulin}
\author[Virginia]{S.~Manecki}
\author[Heidelberg]{W.~Maneschg}
\author[Genova]{G.~Manuzio}
\author[Munich]{Q.~Meindl}
\author[Milano]{E.~Meroni}
\author[Milano]{L.~Miramonti}
\author[Krakow]{M.~Misiaszek}
\author[LNGS,Princeton]{D.~Montanari}
\author[Peters]{V.~Muratova}
\author[Munich]{L.~Oberauer}
\author[APC]{M.~Obolensky}
\author[Perugia]{F.~Ortica}
\author[Genova]{M.~Pallavicini}
\author[LNGS]{L.~Papp}
\author[Milano]{L.~Perasso}
\author[Genova]{S.~Perasso}
\author[UMass]{A.~Pocar}
\author[Virginia]{R.S.~Raghavan}
\author[Milano]{G.~Ranucci}
\author[LNGS]{A.~Razeto}
\author[Milano]{A.~Re}
\author[Ferrara]{B.~Ricci}
\author[Genova]{P.~Risso}
\author[Perugia]{A.~Romani}
\author[Virginia]{D.~Rountree}
\author[Kurchatov]{A.~Sabelnikov}
\author[Princeton]{R.~Saldanha}
\author[Genova]{C.~Salvo}
\author[Heidelberg]{S.~Sch\"onert}
\author[Heidelberg]{H.~Simgen}
\author[Kurchatov]{M.~Skorokhvatov}
\author[Dubna]{O.~Smirnov}
\author[Dubna]{A.~Sotnikov}
\author[Kurchatov]{S.~Sukhotin}
\author[LNGS]{Y.~Suvorov}
\author[LNGS]{R.~Tartaglia}
\author[Genova]{G.~Testera}
\author[APC]{D.~Vignaud}
\author[Virginia]{R.B.~Vogelaar}
\author[Munich]{F.~von~Feilitzsch}
\author[Munich]{J.~Winter}
\author[Krakow]{M.~Wojcik}
\author[Princeton]{A. Wright}
\author[Munich]{M.~Wurm}
\author[Princeton]{J.~Xu}
\author[Dubna]{O.~Zaimidoroga}
\author[Genova]{S.~Zavatarelli}
\author[Heidelberg]{G.~Zuzel}
\author{\\(Borexino Collaboration)}
\address[APC]{Laboratoire AstroParticule et Cosmologie, 75231 Paris cedex 13, France}
\address[Dubna]{Joint Institute for Nuclear Research, 141980 Dubna, Russia}
\address[Genova]{Dipartimento di Fisica, Universit\`a e INFN, Genova 16146, Italy}
\address[Heidelberg]{Max-Planck-Institut f\"ur Kernphysik, 69029 Heidelberg, Germany}
\address[Krakow]{M.~Smoluchowski Institute of Physics, Jagiellonian University, 30059 Krakow, Poland}
\address[Kurchatov]{RRC Kurchatov Institute, 123182 Moscow, Russia}
\address[LNGS]{INFN Laboratori Nazionali del Gran Sasso, SS 17 bis Km 18+910, 67010 Assergi (AQ), Italy}
\address[Milano]{Dipartimento di Fisica, Universit\`a degli Studi e INFN, 20133 Milano, Italy}
\address[Munich]{Physik Department, Technische Universit\"at Muenchen, 85747 Garching, Germany}
\address[Perugia]{Dipartimento di Chimica, Universit\`a e INFN, 06123 Perugia, Italy}
\address[Peters]{St. Petersburg Nuclear Physics Institute, 188350 Gatchina, Russia}
\address[Princeton]{Physics Department, Princeton University, Princeton, NJ 08544, USA}
\address[PrincetonChemEng]{Chemical Engineering Department, Princeton University, Princeton, NJ 08544, USA}
\address[UMass]{Physics Department, University of Massachusetts, Amherst, MA 01003, USA}
\address[Virginia]{Physics Department, Virginia Polytechnic Institute and State University, Blacksburg, VA 24061, USA}
\address[Ferrara]{Dipartimento di Fisica, Universit\`a degli Studi di Ferrara, Ferrara I-44100, Italy}
\address[Kiev]{Institute for Nuclear Research, 03680 Kiev, Ukraine}
\begin{abstract}

Geo--neutrinos, electron anti--neutrinos produced in $\beta$ decays of naturally occurring radioactive isotopes in the Earth, are a unique direct probe of our planet's interior.  
We report the first observation at more than 3$\sigma$ C.L.  of geo--neutrinos, performed with the Borexino detector at Laboratori Nazionali del Gran Sasso.  Anti--neutrinos are detected through the neutron inverse $\beta$ decay reaction.  With a \exposure\ fiducial exposure after all selection cuts, we detected $9.9^{+4.1}_{-3.4}(^{+14.6}_{-8.2})$ geo--neutrino events, with errors corresponding to a 68.3\%~(99.73\%)~C.L.  From the $\ln{\cal{L}}$ profile, the statistical significance of the Borexino geo-neutrino observation corresponds to a 99.997\%~C.L.

Our measurement of the geo--neutrinos rate is $3.9^{+1.6}_{-1.3}(^{+5.8}_{-3.2})$\,\unit.  

The observed prompt positron spectrum above 2.6\,MeV is compatible with that expected from european nuclear reactors (mean base line of approximately 1000\,km).  Our measurement of reactor anti--neutrinos excludes the non-oscillation hypothesis at 99.60\%~C.L.
This measurement rejects the hypothesis of an active geo-reactor in the Earth's core with a power above 3\,TW at 95\%~C.L.

\end{abstract}
\end{frontmatter}
Geo--neutrinos (geo--$\bar{\nu}_{e}$) are electron anti--neutrinos ($\bar{\nu}_{e}$) produced in $\beta$ decays of $^{40}$K and of several nuclides in the chains of long--lived radioactive isotopes $^{238}$U and $^{232}$Th, which are naturally present in the Earth.  Information about the Earth's interior composition has insofar come exclusively from indirect probes: seismology only constrains the density profile, while geochemistry offers previsions based on chemical compositions of rocks from the upper Earth layers, chondritic meteorites, and the photosphere of the Sun.  Geo--$\bar{\nu}_{e}$'s are direct messengers of the abundances and distribution of radioactive elements within our planet.  By measuring their flux and spectrum it is possible to reveal the distribution of long-lived radioactivity in the Earth and to assess the radiogenic contribution to the total heat balance of the Earth.  These pieces of information, in turn, are critical in understanding complex processes such as the generation of the Earth's magnetic field, mantle convection, and plate tectonics.

Geo--$\bar{\nu}_{e}$'s were introduced by Eder~\cite{Eder} and Marx~\cite{Marx} in the 1960's.  The subject was thoroughly reviewed by Krauss et al. in 1984~\cite{Krauss}.  The possibility of detecting geo--neutrinos with large scintillator-based solar neutrino detectors was pointed out in Refs.~\cite{Raghavan,Calaprice,CTFAntiNu,Fiorentini}.  A first experimental indication for geo--$\bar{\nu}_{e}$ ($\sim$2.5~$\sigma$~C.L.) was reported by the KamLAND collaboration~\cite{KamLAND1, KamLAND2}.

This letter reports the first observation of geo--$\bar{\nu}_{e}$, performed with the Borexino detector at Laboratori Nazionali del Gran Sasso (LNGS).  In the context, we also measured $\bar{\nu}_{e}$ from distant nuclear reactors with a mean--base line of approximately 1000 km, as first discussed in~\cite{Schoenert}.

Borexino is an unsegmented liquid scintillator detector built for the observation and measurement of low-energy solar neutrinos.  The Borexino Collaboration already reported the observation of $^7$Be solar--$\nu_e$~~\cite{BX7BePLB,BX7BePRL} and the measurement of $^8$B solar--$\nu_e$~\cite{BX8B}.

The liquid scintillator consists of 278\,tons of pseudocumene (PC) doped with 1.5~g/l of diphenyloxazole (PPO), confined within a thin spherical nylon vessel with a radius of 4.25\,m.  It is shielded from external radiation by 890 tons of liquid buffer, a solution of PC and 5.0~g/l of the light quencher dimethylphthalate (DMP).  A second spherical nylon vessel with a 5.75~m radius segments the liquid buffer in two contiguous volumes and prevents diffusion of the radon emanating from the periphery of the detector close to the liquid scintillator.  The liquid buffer is contained in a 13.7~m diameter stainless steel sphere (SSS).  The SSS is housed in a 9\,m-radius, 16.9\,m high domed water tank (WT), filled with ultra-high purity water, which serves as a passive shield against neutrons and gamma-rays.  Scintillation light is detected by 2212~8''~PMTs (the Inner Detector, ID).  \v{C}erenkov light radiated by muons passing through the water is measured by 208~8" external PMTs (the Outer Detector, OD).  A detailed description of the Borexino detector can be found in Ref.~\cite{BXdetector,BXfluid}. 

The unprecedentedly low intrinsic radioactivity achieved in Borexino, the high photon yield, and the large number of free target protons ($\sim$1.7$\times 10^{31}$) offer a unique opportunity for a sensitive search for $\bar{\nu}_e$'s in the~MeV energy range.

Borexino detects $\bar{\nu}_{e}$ via the inverse neutron $\beta$ decay,
\begin{equation}
\bar{\nu}_e + p \rightarrow e^+ + n,
\label{Eq:InvBeta}
\end{equation}
with a threshold of 1.806\,MeV.  Some $\bar{\nu}_{e}$ from the $^{238}$U and $^{232}$Th series are above threshold, while those from $^{40}$K decays are below threshold.  The positron from the inverse neutron $\beta$ decay promptly comes to rest in the liquid scintillator and annihilates emitting two 511\,keV $\gamma$-rays, yielding a prompt event, with a visible energy of $E_{\rm prompt} = E_{\bar{\nu}_e} - 0.782\,{\rm MeV}$ (the scintillation light related to the proton recoil is highly quenched and negligible).  The free neutron emitted is typically captured on protons with a mean time of $\tau$$\sim$256\,$\mu$s, resulting in the emission of a 2.22\,MeV de-excitation $\gamma$-ray, which provides a coincident delayed event.  The characteristic time and spatial coincidence of prompt and delayed events offers a clean and unmistakable signature of $\bar{\nu}_e$ detection.

In this paper we report the analysis of data collected between December 2007~and December~2009, corresponding to 537.2\,days of live time.  The fiducial exposure after cuts is \exposure.

\begin{figure}[t!]
\includegraphics[scale=0.42,angle=90]{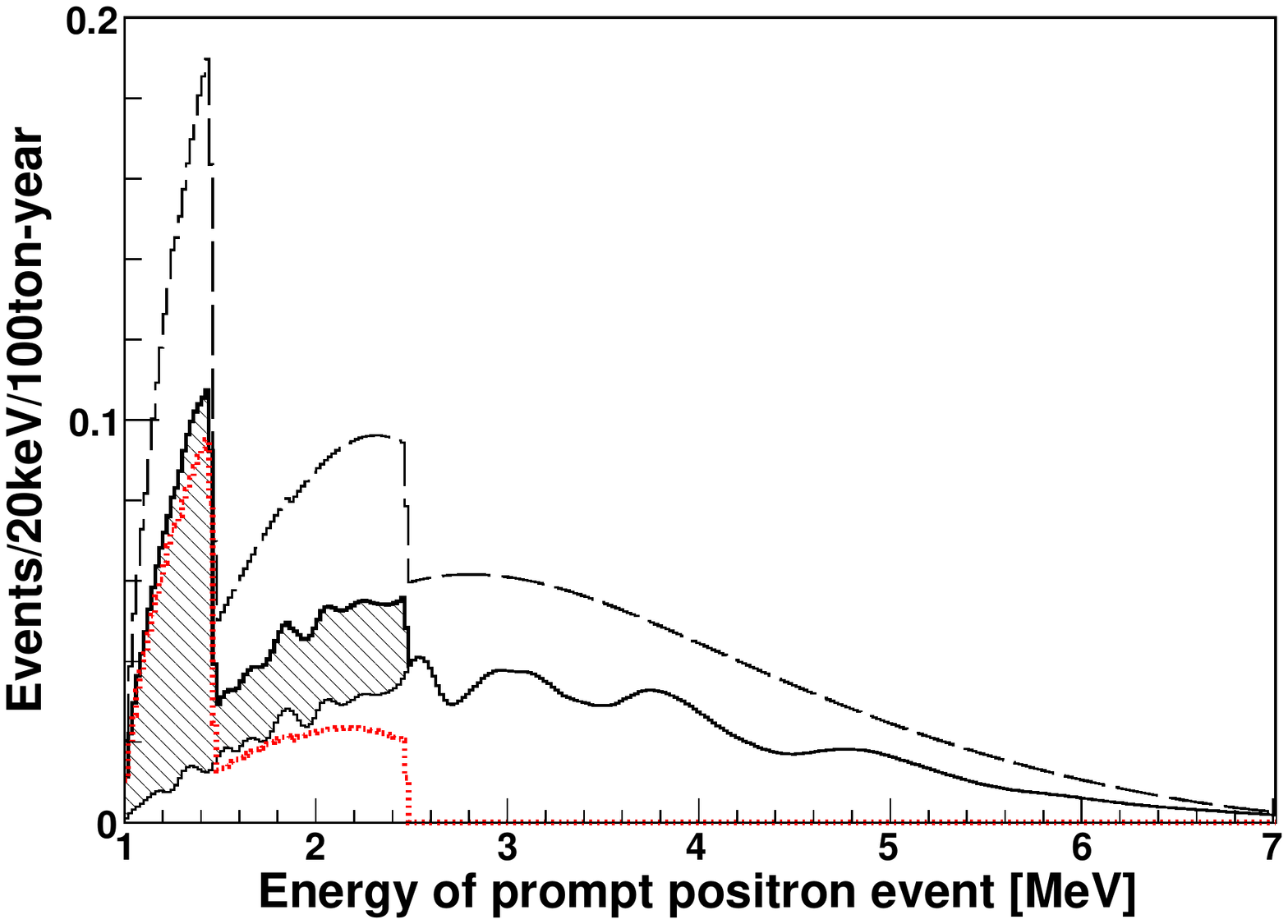}
\caption{Expected spectrum for electron anti-neutrinos in Borexino.  The horizontal axis shows the kinetic plus the annihilation 1.022 MeV energy of the prompt positron event. Dashed line: total geo--$\bar{\nu}_e$ plus reactor--$\bar{\nu}_e$ spectrum without oscillations. Solid thick lines: geo--$\bar{\nu}_e$ and reactor--$\bar{\nu}_e$ with oscillations.  Dotted line (red): geo--$\bar{\nu}_e$ with the high (low) energy peak due to decays in the $^{238}$U chain ($^{238}$U+$^{232}$Th chains).  Solid thin line: reactor--$\bar{\nu}_e$. See text for details.}
\label{Fig:spcEnergy}
\end{figure}

Fig.~\ref{Fig:spcEnergy} shows the expected $E_{\rm prompt}$ spectrum.  It includes signals from geo--$\bar{\nu}_e$ (up to $\sim$2.6\,MeV) and reactor $\bar{\nu}_e$ (up to $\sim$8\,MeV).  For geo--$\bar{\nu}_e$'s we have used known energy spectra of $\beta^{-}$ decays, the chondritic Th/U mass ratio of 3.9, and the geo--$\bar{\nu}_e$ fluxes obtained from the Bulk Silicate Earth (BSE) geochemical model~\cite{Mantovani}, which predicts a detection rate of 2.5$^{+0.3}_{-0.5}$ events/(100\,ton$\cdot$yr) for geo--neutrinos in Borexino.

The determination of the expected signal from reactor $\bar{\nu}_e$'s required the collection of the detailed information on the time profiles of power and nuclear fuel composition for nearby reactors.  The differential reactor anti-neutrino spectrum, in units of ${\bar{\nu}_e}$/(MeV cm$^2$), is:
\begin{equation}
\begin{split}
\Phi (E_{\bar{\nu}_e}) = & \sum_{r=1}^{N_{react}} \sum_{m=1}^{N_{month}} \frac {T_{m}}{4 \pi L_{r}^{2}} P_{rm}  \times   \\
\times & \sum_{i=1}^4 \frac {f_{i}}{E_{i}} \phi_{i}(E_{\bar{\nu}_e}) P_{ee}(E_{\bar{\nu}_e};\hat\theta, L_r)
\end{split}
\label{Eq:ReactorFlux}
\end{equation}
where the index $r$ cycles over the $N$ reactors considered, the index $m$ cycles over the total number of months $M$ for the present data set, $T_m$ is the live time in the $m^{\rm th}$ month, $L_{r}$ is the distance of the detector from reactor $r$, $P_{rm}$ is the effective thermal power of reactor $r$ in month $m$, the index $i$ stands for the i-th spectral component in the set ($^{235}$U, $^{238}$U, $^{239}$Pu, and $^{241}$Pu), $f_{i}$ is the power fraction of the component $i$, $E_i$ is the average anti--neutrino energy per fission of the component $i$, $\phi(E_{\bar{\nu}})$ is the anti-neutrino flux per fission of the $i^{\rm th}$ component, and $P_{ee}$ is the survival probability of the reactor anti-neutrinos of energy $E_{\bar{\nu}}$ traveling the baseline $L_r$, for mixing parameters $\hat\theta$ = ($\Delta m_{12}^2$, $\sin\theta^2_{12}$).  In Eq.~(\ref{Eq:ReactorFlux}) the main contribution comes from 194 reactors in Europe, while other 245 reactors around the world~\cite{WorldReactors} contribute only 2.5\% of the total reactor signal.  The ${\bar{\nu}_e}$ energy spectra, $\phi_i(E_{\bar{\nu}_e})$ in Eq.~(\ref{Eq:ReactorFlux}), are taken from~\cite{Huber}.  Typical power fractions for the fuel components are:
\begin{equation}
\begin{split}
\mbox{$^{235}$U} & : \mbox{$^{238}$U} : \mbox{$^{239}$Pu}  : \mbox{$^{241}$Pu}   =  \\
0.56 & :  \,0.08 \,:  \,0.30 \, \, \, : \, 0.06  \\
 \end{split}
\end{equation}
with a systematic error of 3.2\% due to possible differences among the fuels of different cores and the unknown stage of burn-up in each reactor.  For the thirty-five european reactors using MOX (Mixed OXide) technology, 30\% of their thermal power was considered to have power fractions:
\begin{equation}
\begin{split}
\mbox{$^{235}$U} & :\mbox{$^{238}$U}:\mbox{$^{239}$Pu}:\mbox{$^{241}$Pu}  = \\
0.000 & :0.080:0.708:0.212  \\
\end{split}
\end{equation}

Information on the nominal thermal power and monthly load factor for each european reactor originates from IAEA and EDF~\cite{IAEA}.

We use the interaction cross section $\sigma_{\bar{\nu}p}$ for inverse--beta decay reaction in Eq.~(\ref{Eq:InvBeta}) from Ref.~\cite{Strumia} and the neutrino oscillations parameters ($\Delta m^2_{12}$ = 7.65~$\cdot 10^{-5}$ eV$^2$; $\sin^2\theta_{12}$ = 0.304) from Ref.~\cite{Valle} (this analysis includes a +0.6\% contribution from matter effects in the approximation of constant Earth density).  The contribution of long--lived fission products in the spent fuel (mainly $^{106}$Ru and $^{144}$Ce) amounts to 1.5\%~\cite{Kopeikin}.  The expected reactor signal with (without) neutrino oscillations and 100\% detection efficiency is 5.7$\pm$0.3\,events/(100\,ton$\cdot$yr) (9.9$\pm$0.5\,events/(100\,ton$\cdot$yr)).  The contributions to the estimated systematic error are summarized in Tab.~\ref{Tab:ReactorError}.  We included a 2\% systematic error arising from the comparison between the IAEA and EDF reactor thermal power data.  A conservative 0.4\% systematic error on distances originates from the uncertainty on the Earth radius, and on the position of the reactor cores and the detector.

\begin{table}[t!]
\caption{Systematic uncertainties on the  expected reactor--$\bar{\nu}_e$ signal.  See Eq.~(\ref{Eq:ReactorFlux}) and accompanying text for details.}
\begin{center}
\begin{tabular}{lrlr}
\hline\hline
Source								&Error	&Source					&Error \\
									&[\%]		&						&[\%] \\ 
\hline
Fuel composition						&3.2		&$\theta_{12}$				&2.6 \\
$\phi(E_{\bar{\nu}})$						&2.5		&$P_{rm}$				&2.0 \\
Long-lived isotopes						&1.0		&$E_i$					&0.6 \\
$\sigma_{\bar{\nu}p}$					&0.4		&$L_{r}$					&0.4 \\
$\Delta m_{12}^2$						&0.02	&						& \\
\hline
Total									&		&						&5.38 \\
\hline\hline
\end{tabular}
\label{Tab:ReactorError} 
\end{center}
\end{table}

An extensive calibration campaign with radioactive sources has been performed in Borexino.  
In October 2008 an on--axis calibration system was used to place $\gamma$, $\beta$, and $\alpha$ sources within the active volume along the vertical axis. In January, February, June, and July 2009 other campaigns were carried out with an off--axis calibration system.  These campaigns included AmBe, $^{57}$Co,  $^{139}$Ce, $^{208}$Hg,  $^{85}$Sr, $^{54}$Mn,  $^{65}$Zn, $^{40}$K, $^{60}$Co, and  $^{222}$Rn.  The AmBe source producing $\sim$10~neutrons/s with energies up to 10\,MeV was deployed in twenty-five different positions allowing the study of the detector response to captured neutrons and to protons recoiling off neutrons.  The measured light yields for gamma-rays following neutron capture on $^1$H (2.22\,MeV) and $^{12}$C (4.95\,MeV) at the center of the detector are 1060$\pm$5 photoelectrons (p.e.) ($\sigma$=42.1$\pm$0.2\,p.e.) and 2368$\pm$20\,p.e. ($\sigma$=72$\pm$3\,p.e.), respectively.

The stability of the detector response is continually monitored during data taking and offline by means of data validation tools.  Muons crossing the liquid scintillator can produce neutrons which are thermalized and captured on $^1H$. The mean light yield produced by the 2.22\,MeV gamma-ray following neutron capture is found stable within 1\%. In the liquid scintillator there is a measurable activity due to $^{210}$Po $\alpha$ decays. The mean light yield produced by the 5.3\,MeV $\alpha$ is measured to be stable within 0.5\%.  The stability of the overall detection efficiency is studied using the measured rates of cosmogenic backgrounds. We clearly see the $\pm$2\% seasonal variation of the muon flux~\cite{MACRO}. For further details on the detector monitoring and the methods of online calibrations  see~\cite{BXdetector}.

\begin{figure}[t!]
\includegraphics[scale=0.42,angle=90]{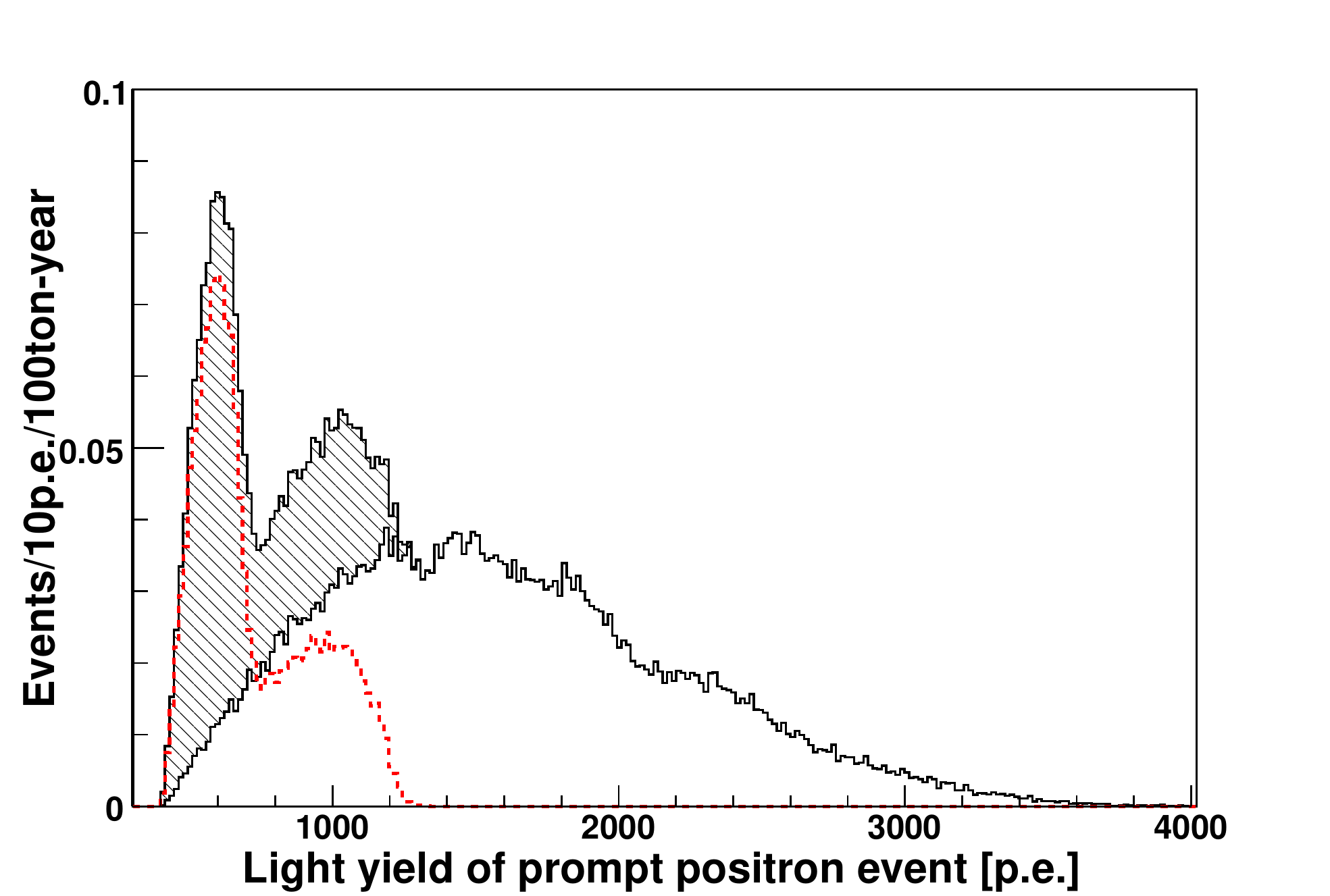}
\caption{Expected prompt positron event spectrum as obtained from the MC code, using the distribution in Fig.~\ref{Fig:spcEnergy} as input and the selection cuts described in the text.  The horizontal axis shows the number of p.e. detected by the PMTs. Primaries generated are $10^5$ events for both geo--$\bar{\nu}_e$ and reactor--$\bar{\nu}_e$.  See text for details.}
\label{Fig:spcMC}
\end{figure}

The {\tt Geant4}-based Borexino Monte Carlo (MC) was tuned on data from the calibration campaign.  The expected geo and reactor $\bar{\nu}_e$'s spectra shown in Fig.~\ref{Fig:spcEnergy} were used as input to the MC code in order to simulate the detector response to $\bar{\nu}_e$ interactions.  The MC--generated geo--$\bar{\nu}_e$ and reactor--$\bar{\nu}_e$ spectra are shown in Fig.~\ref{Fig:spcMC}, where the energy is expressed as the total light yield (in units of p.e.) collected by the PMTs.

In Borexino, the position of each event is determined from the timing pattern of hit PMTs.  The Fiducial Volume (FV) is determined with 3.8\% uncertainty, based on the source calibration campaign.  The maximal deviation from the calibration reference positions measured at 4\,m radius is 5\,cm.

The event energy is a calibrated non--linear function of the number of detected p.e.  The total number of p.e. collected by the PMTs, $Q$, depends on the energy, position, and nature of the events (light yield of $\gamma$-rays and electrons differ slightly due to the light quenching of low energy electrons).  All these dependences are properly handled and well reproduced by the MC code, permitting us to perform the analysis directly on the light yield spectrum shown in Fig.~\ref{Fig:spcMC} (rather than on the energy spectrum of Fig.~\ref{Fig:spcEnergy}).

The following cuts are used for $\bar{\nu}_e$'s search: $Q_{\rm prompt}$$>$410\,p.e., where $Q_{\rm prompt}$ is the PMTs light yield for the prompt event; 700\,p.e.$<$$Q_{\rm delayed}$$<$1,250\,p.e., where $Q_{\rm delayed}$ is the PMTs light yield for the delayed event; $\Delta R$$<$1\,m, where $\Delta R$ is the reconstructed distance between the prompt and the delayed event; 20\,$\mu$s$<$$\Delta t$$<$1280\,$\mu$s, where $\Delta t$ is the time interval between the prompt and the delayed event.
The selection criterium for the reconstructed radius of the prompt event,  $R_{\rm prompt}$, sets a fiducial volume, providing a 0.25\,m layer of active shielding against external backgrounds.  The total detection efficiency with these cuts was determined by MC to be 0.85$\pm$0.01.

An important source of background to the $\bar{\nu}_e$'s measurement is due to $\beta^-$--neutron emitters produced by muons interacting in the scintillator, {\it i.e.} $^9$Li ($\tau$=260\,ms) and $^8$He ($\tau$=173\,ms)~\cite{Hagner}.  We reject these events by applying a 2\,s veto after each muon crossing the liquid scintillator active volume.  The veto inefficiency on the background from $\beta^-$--neutron emitters is $3\cdot10^{-5}$.  We tagged fifty-one (51) $^9$Li--$^8$He candidates falling within the $\bar{\nu}_e$ cuts in coincidence with a positive signal from the above veto, corresponding to a measured $^9$Li--$^8$He rate of 15.4\,\unit.  The residual $^9$Li--$^8$He background after the muon cut is equal to 0.03$\pm$0.02\,\unit.

\begin{table}[t!]
\caption{Estimated backgrounds for the $\bar{\nu}_e$'s.  Upper limits are given at 90\%~C.L.}
\begin{center}
\begin{tabular}{lc}
\hline\hline
Source						&Background \\
							&[\unit] \\
\hline
$^9$Li--$^8$He				&0.03$\pm$0.02 \\ 
Fast $n$'s ($\mu$'s in WT)		&$<$0.01 \\
Fast $n$'s ($\mu$'s in rock)		&$<$0.04 \\
Untagged muons				&0.011$\pm$0.001 \\
Accidental coincidences			&0.080$\pm$0.001 \\
Time corr. background			&$<$0.026 \\
($\gamma$,n)					&$<$0.003  \\
Spontaneous fission in PMTs				&0.0030$\pm$0.0003 \\
($\alpha$,n) in scintillator			&0.014$\pm$0.001 \\
($\alpha$,n) in the buffer		&$<$0.061\\
\hline
Total							&0.14$\pm$0.02  \\ 
\hline\hline
\end{tabular}
\label{Tab:Bckg} 
\end{center}
\end{table}

Fast neutrons can mimic $\bar{\nu}_e$ events: recoiling protons scattered by the neutron during its thermalization can fake a prompt signal, and the thermalized neutron capture on a proton produces a 2.22\,MeV $\gamma$--ray delayed signal.  Fast neutrons contributing to our background can be produced by muons either crossing the Borexino WT or interacting in the rock around the detector.

We reject more than 99.5\% of fast neutrons originated within the WT with a 2\,ms veto following each muon crossing the WT but not the SSS.  We have identified two (2) candidates faking a $\bar{\nu}_e$ event in coincidence with muons crossing the WT.  Thus, the background from undetected muons crossing the WT is estimated as $<$0.01\,\unit\ with a 90\%~C.L.

Fast neutrons originated by muons in the rocks surrounding the detector have an average energy of $\langle E \rangle$$\sim$90\,MeV and can penetrate for a few meters inside the detector, and eventually reach the active scintillator target.  The background rate from these events was studied with a MC simulation, which used as input the energy spectrum of fast neutrons reported in Ref.~\cite{Hime} for the specific case of LNGS.  We estimate this background at $<$0.04\,\unit\ with a 90\%~C.L.

Muons are typically identified and rejected by the OD, but can also be distinguished from a point--like scintillation event by the pulse shape analysis of the ID signal.  Several categories of muons which might have gone undetected or not identified as muons, were studied in detail and the possibility that they cause a false $\bar{\nu}_e$ event was considered.

First, primary muons can mimic the prompt signal and a muon--induced neutron a delayed signal.  This background is strongly suppressed, since muons depositing a visible energy below 8\,MeV (our range of interest, see Fig.~\ref{Fig:spcEnergy}), cross the WT and the buffer region without entering the scintillator: the probability that the muon-induced neutron is detected and falls within our cuts is negligible.  Out of  0.7 millions detected buffer muons, eighteen (18) $\bar{\nu}_e$ candidates were selected, $2.5\cdot10^{-5}$ per muon.

Second, pairs of muon--induced neutrons following unrecognized muons can also simulate $\bar{\nu}_e$ events.  Most muons producing more than one detected neutron have crossed the scintillator and are hence tagged with very high efficiency.  In addition, many muon--induced neutrons are produced with multiplicity higher than two.  The requirement that, within a 2\,ms window, every $\bar{\nu}_e$ candidate is neither preceded nor followed by another event with neutron-like energy strongly suppresses this background.

The estimated combined background from muon-neutron and neutron-neutron coincidences is 0.011$\pm$0.001\,\unit.

We determined the background from accidental coincidences by using an off--time coincidence window of 2--20\,s.  By simple scaling, in the 1260~$\mu$s wide time window used in the $\bar{\nu}_e$ search, the number of accidental coincidences is 0.080$\pm$0.001\,\unit.

To ensure the absence of any unknown time-correlated background we performed a detailed study using an off--time coincidence window of 2\,ms--2\,s.  We conclude that no significant time-correlated events are included in the data sample, with a limit of $<$0.026\,\unit\ at 90\%~C.L.

Radioactivity in the nylon vessels, in the PMTs and, in the stainless steel can induce ($\gamma$,n) and spontaneous fission reactions which produce MeV neutrons and hence mimic $\bar{\nu}_e$'s events.  On the basis of the known radioactivity of the components, we estimate these backgrounds at $<$0.003 (90\%~C.L.) and 0.0030$\pm$0.0003, respectively.

Neutrons of energies up to 7.3 MeV can also arise from $^{13}$C($\alpha$,$n$)$^{16}$O reactions following $^{210}$Po $\alpha$ decays, as investigated by KamLAND~\cite{KamLAND2}.  $^{210}$Po is the main contaminant in Borexino~\cite{BX7BePRL}, with a decay rate of 12$\pm$1\,counts/(ton$\cdot$day) on average for the present data set.  Yet its abundance is many orders of magnitude lower than in the KamLAND scintillator.  We estimate the probability for a $^{210}$Po $\alpha$ to trigger an ($\alpha$,$n$) reaction as (5.0$\pm$0.3)$\times$10$^{-8}$, from data from Ref.~\cite{McKee}.  This source of background, which is quantitatively very important in the $\bar{\nu}_e$'s search in KamLAND, yields an almost negligible 0.014$\pm$0.001\,\unit\ in Borexino, thanks to the much lower level of intrinsic $^{210}$Po background.

Another source of background are ($\alpha$,n) reactions due to $^{210}$Po decays in the buffer.  $^{210}$Po contamination in the buffer was measured by counting $\alpha$'s from a sample of buffer fluid prepared in a vial and lowered at the center of the Counting Test Facility of Borexino~\cite{CTF}.  We obtained an upper limit of $<$~0.67\,mBq/kg at 90\%~C.L. for the contamination from $^{210}$Po in the buffer, an upper limit which is many orders of magnitude above the measured $^{210}$Po contamination in the scintillator.  A shallow radial cut is very effective in removing possible background originating from the buffer, due to the short distance of thermalization of the low-energy ($\alpha$,$n$) neutrons.  We estimate via MC simulations that the radial cut reduces 14-fold the background induced by $^{210}$Po in the buffer, corresponding to an upper limit for this background of $<$~0.061 \,\unit\  at 90\%~C.L.

Table~\ref{Tab:Bckg} summarizes all expected backgrounds obtained by scaling from  \exposure\ fiducial exposure to 100 ton$\cdot$yr for the sake of clarity.  Independent errors are summed in quadrature.  In conclusion, we expect 0.14$\pm$0.02\,\unit\,  background events in the Borexino search for $\bar{\nu}_e$'s.

\begin{figure}[t!]
\includegraphics[scale=0.42,angle=90]{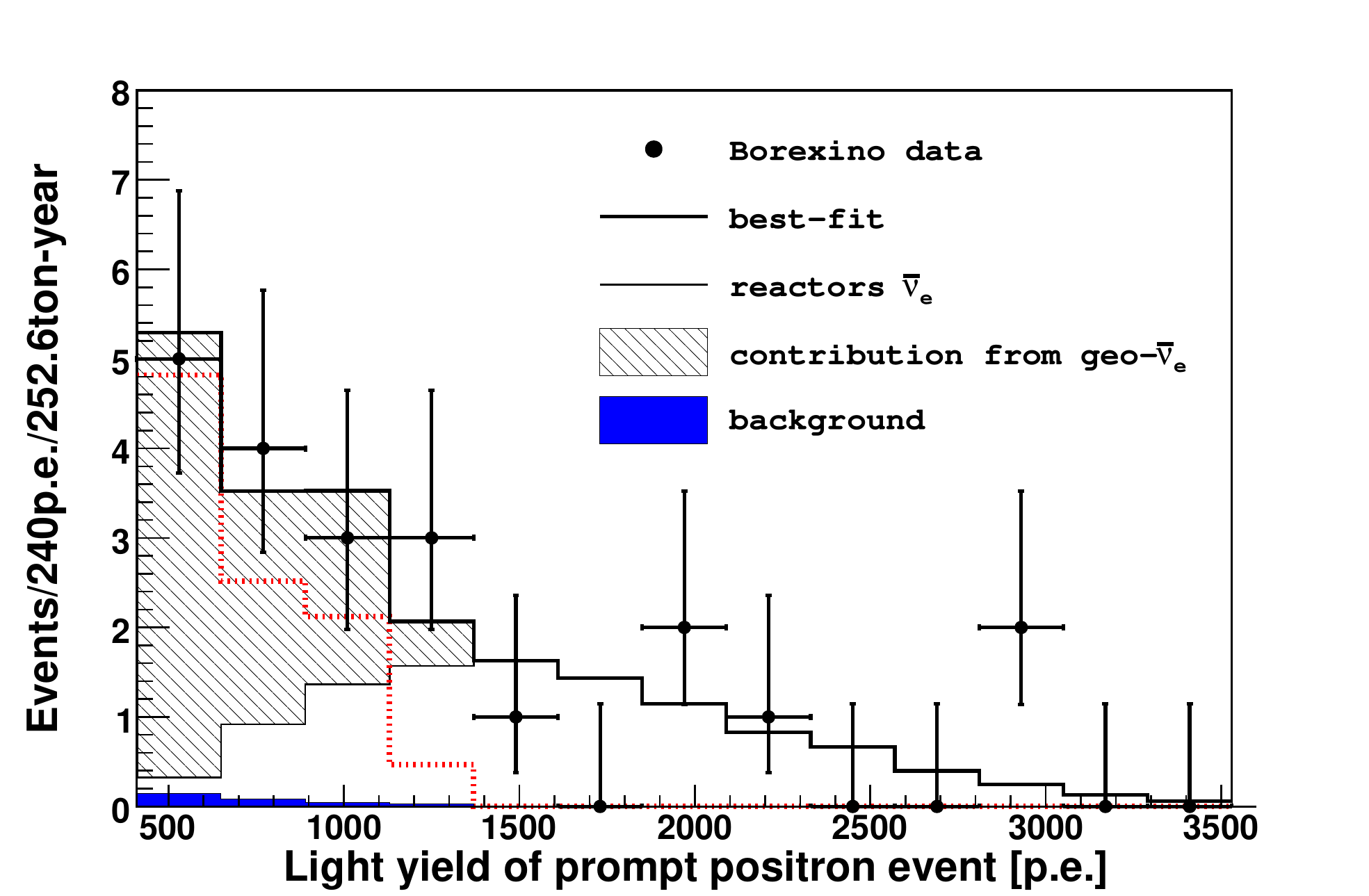}
\caption{Light yield spectrum for the positron prompt events of the 21 $\bar{\nu}_e$ candidates and the best-fit with Eq.~(\ref{Eq:fit}) (solid thick line).  The horizonal axis shows the number of p.e. detected by the PMTs. The small filled area on the lower left part of the spectrum is the background.  Thin solid line: reactor--$\bar{\nu}_e$ signal from the fit.  Dotted line (red): geo--$\bar{\nu}_e$ signal resulting from the fit.  The darker area isolates the contribution of the geo--$\bar{\nu}_e$ in the total signal. The conversion from p.e. to energy is approximately 500 p.e./MeV.}
\label{Fig:data}
\end{figure}

A total of twenty-one (21) $\bar{\nu}_e$'s candidates pass all selection cuts described above.  
The spatial and time distributions of the candidates is uniform within the limited statistics of the observed sample.
The expected number of background events, in the present data set is 0.40$\pm$0.05.  The signal to background ratio in the $\bar{\nu}_e$'s Borexino search is an unprecedented $\sim$50:1.

As shown by our MC (see Fig.~\ref{Fig:spcMC}), the light yield spectrum of the prompt events below 1,300\,p.e. includes 100\% of the geo--$\bar{\nu}_e$ signal and only 34.7\% of the reactor--$\bar{\nu}_e$ signal.  We do not know any other source of $\bar{\nu}_e$'s which could give a considerable contribution in this region. We notice that atmospheric and supernova relic $\bar{\nu}_e$'s give a negligible contribution.
No geo--$\bar{\nu}_e$ are expected above 1,300\,p.e. 
We note from Fig.~\ref{Fig:data} that of the total of twenty-one (21) candidates, fifteen (15) are in the {geo--$\bar{\nu}_e$ energy window below 1,300 p.e. and six (6) have a light yield exceeding 1,300\,p.e.  The 50:1 signal to background ratio and the clear separation of the two $\bar{\nu}_e$ sources in the light yield spectrum of the prompt event permit a clear identification and separation of the number of events belonging to each source, and allow to establish observation of the geo--neutrinos, as described below.

Experimental evidence of reactors $\bar{\nu}_e$ disappearance and oscillations has been reported by the KamLAND collaboration on a mean base line of approximately 200 km~\cite{KamLANDosc,KamLAND2}.  Based on our calculation of anti--$\bar{\nu}_e$ fluxes from reactors, in the reactor--$\bar{\nu}_e$ window (Q$_{\rm prompt}$$>$1,300\,p.e.) we expect 16.3$\pm$1.1\,events in absence of neutrino oscillations and 9.4$\pm$0.6\,events in presence of neutrino oscillations with parameters as determined in Ref.~\cite{Valle}.  The expected background in the reactor--$\bar{\nu}_e$ window is 0.09$\pm$0.06.  We observe in the reactor--$\bar{\nu}_e$ window six (6) events, and we conclude that our measurement of reactor--$\bar{\nu}_e$ is statistically compatible with the expected signal in presence of neutrino oscillation.  A statistical analysis excludes the no-oscillation hypothesis at 99.60\%~C.L.

In the geo--$\bar{\nu}_e$ window (Q$_{\rm prompt}$$<$1,300\,p.e.) we expect 5.0$\pm$0.3\,events from reactors (under the hypothesis of oscillations with the mixing parameters quoted above) and 0.31$\pm$0.05 background events.  We observe in the geo--$\bar{\nu}_e$  window fifteen (15) candidates.

The hypothesis that the excess of events is due to a statistical fluctuation of the background plus the reactor events is rejected at the 99.95\%~C.L.

\begin{figure}[t!]
\centerline{\includegraphics[width= 7 cm]{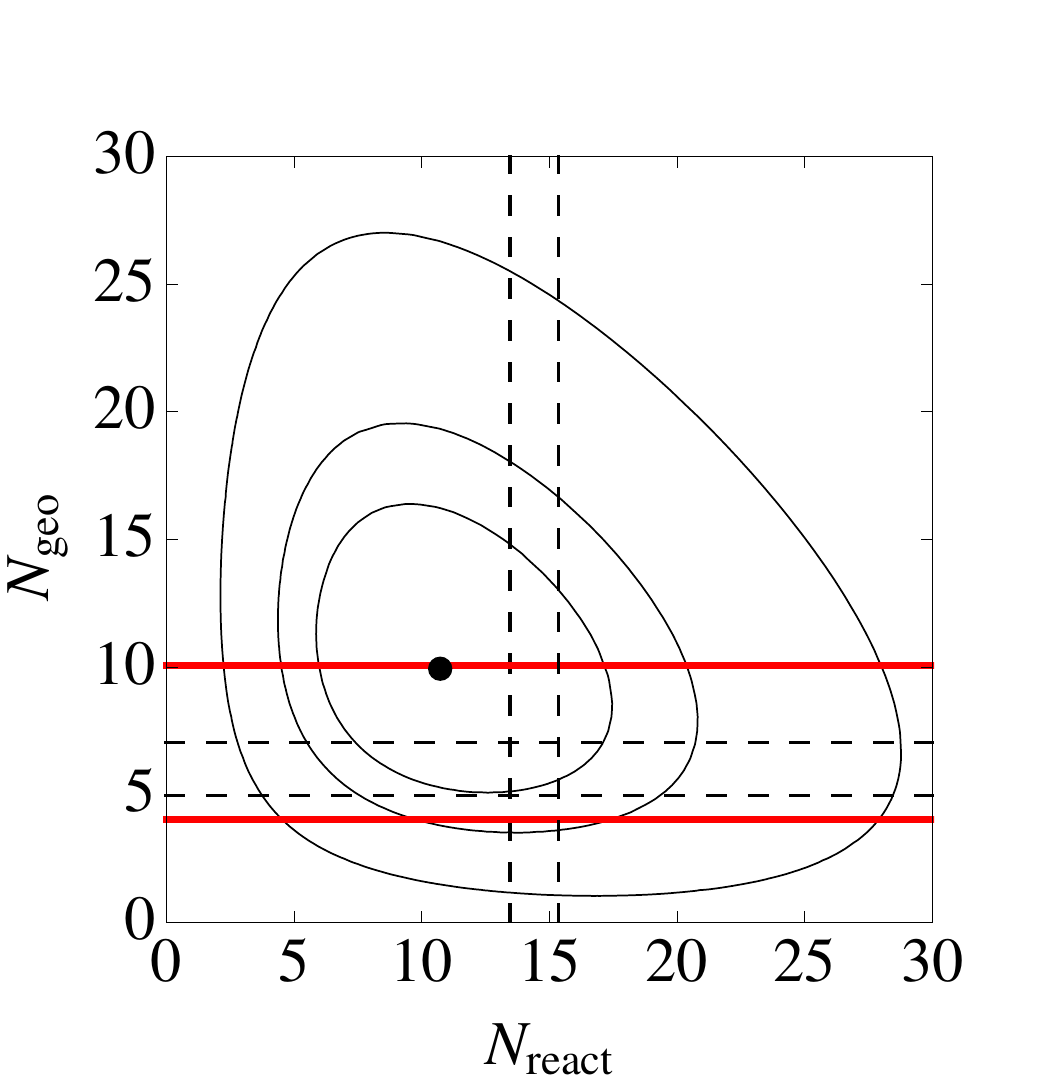}}
\caption{\label{Fig:contour} Allowed regions for $N_{\rm geo}$ and $N_{\rm react}$ at 68\%, 90\%, and 99.73\%~C.L.  Vertical dashed lines: 1$\sigma$ range about the expected $N_{\rm react}$ (expected in presence of neutrino oscillations).  Horizontal dashed lines: range for $N_{\rm geo}$ predictions based on the BSE model in Ref.~\cite{Fiorentini}.  Horizontal solid red lines: predictions of the Maximal and Minimal Radiogenic Earth models.  See text for details.}
\end{figure}

Finally, we determine our best estimate of the geo--$\bar{\nu}_e$ and of the reactor--$\bar{\nu}_e$ rates with an unbinned maximum likelihood analysis of the twenty-one (21) observed $\bar{\nu}_e$ candidates.  Of maximal interest is the light yield of the prompt events of the candidates, which, as shown above, permits to disentangle the two classes of events.  We define the log-likelihood function as~\cite{Cowan,Ianni}: \\
\begin{equation}
\begin{split}
\ln{\cal{L}}&(N_{\rm geo},N_{\rm react},S_{\rm react},S_{\rm FV}) = \\
-& N_{\rm expected}(N_{\rm geo},N_{\rm react},S_{\rm react},S_{\rm FV}) +\\
+& \sum_{i=1}^{N}\ln [f_{\bar{\nu}}(Q_i,N_{\rm geo},N_{\rm react},S_{\rm react},S_{\rm FV})+f_B(Q_i)] \\
-& \frac{1}{2} {\left[ \left(\frac{S_{\rm react}}{\sigma_{\rm react}}\right)^2 + \left(\frac{S_{\rm FV}}{\sigma_{\rm FV}}\right)^2 \right]}    
\end{split}
 \label{Eq:fit}
\end{equation}
where the index $i$ cycles over the $N$=21 candidate events, $Q_i$ is the light yield of the prompt event in p.e. for the $i^{\rm th}$ candidate, N$_{\rm geo}$ and N$_{\rm react}$ are the number of geo--$\bar{\nu}_e$ and of reactor--$\bar{\nu}_e$, the terms $S_{\rm react}$ and $S_{\rm FV}$ account for systematic uncertainties, $\sigma_{\rm react}$=0.0538 and $\sigma_{FV}$=0.038 represent the fractional uncertainties on the reactor neutrino fluxes and on the fiducial volume described earlier in the text, $N_{\rm expexted}$ is the expected total number of $\bar{\nu}_e$, $f_B$ is the spectrum of backgrounds quoted in Tab.~\ref{Tab:Bckg}, $f_{\bar{\nu}}=(1+S_{FV}) [f_{\rm geo}+(1+S_{react}) f_{\rm react}]$ is global $\bar{\nu}_e$ spectrum, and $f_{\rm geo}$ and $f_{\rm react}$ are the individual spectra of the geo--$\bar{\nu}_e$ and of the reactor--$\bar{\nu}_e$, respectively.

Fig.~\ref{Fig:data} shows the comparison of the data with the best likelihood fit.  Our best estimates are $N_{\rm geo}$=$9.9^{+4.1}_{-3.4}(^{+14.6}_{-8.2})$ and $N_{\rm react}$=$10.7^{+4.3}_{-3.4}(^{+15.8}_{-8.0})$ at 68.3\%~C.L. (99.73\%~C.L.).  Fig.~\ref{Fig:contour} shows the allowed regions for $N_{\rm geo}$ and  $N_{\rm react}$.  By studying the profile of the log-likelihood with respect to $N_{\rm geo}$ we have calculated that the null hypothesis for geo--$\bar{\nu}_e$ ({\it i.e.}, $N_{\rm geo}$=0) can be rejected at 99.997\%~C.L., which represents the statistical significance of the observation of geo--$\bar{\nu}_e$ reported in this paper.

Scaling the best estimate of $N_{\rm geo}$ with the \exposure\ exposure, we obtain as our best measurement for the geo--neutrinos rate $3.9^{+1.6}_{-1.3}(^{+5.8}_{-3.2})$\,\unit.  In Tab.~\ref{Tab:Results} we compare the measured rate with predictions of some of the most interesting geophysical models.  In particular, we report as terms of comparison upper and lower bounds on the BSE models, considering the spread of U and Th abundances and their distributions allowed by this geochemical model; the expectation under the Minimal Radiogenic Earth scenario, which considers U and Th from only those Earth layers whose composition can be studied on direct rock--samples; the expectation under the Maximal Radiogenic Earth scenario, which assumes that all terrestrial
heat (deduced from measurements of temperature gradients along $\sim$20,000 drill holes spread over the World) is produced exclusively by radiogenic elements.

\begin{table}[t!]
\caption{Comparison the Borexino measurement of geo--$\bar{\nu}_e$ with predictions.  See text for details.}
\begin{center}
\begin{tabular}{lc}
\hline\hline
Source					&Geo--$\bar{\nu}_e$ Rate \\
						&[\unit] \\
\hline
Borexino					&$3.9^{+1.6}_{-1.3}$ \\
\hline
BSE~\cite{Mantovani}		& $2.5^{+0.3}_{-0.5}$ \\ 
BSE~\cite{Fogli}			&2.5$\pm$0.2 \\
BSE~\cite{Calaprice}		&3.6 \\
Max. Radiogenic Earth		& 3.9 \\ 
Min. Radiogenic Earth		&1.6 \\
\hline\hline
\end{tabular}
\label{Tab:Results}
\end{center}
\end{table}
 
The data presented in this letter unambiguously show, despite the limited statistics, the sensitivity of Borexino for detecting geo-neutrinos.  Thanks to the extraordinarily low background and its unprecedented 50:1 signal to background ratio obtained in the $\bar{\nu}_e$ search, we establish observation of geo--neutrinos at 4.2$\sigma$ C.L. The ratio between the measured geo--$\bar{\nu}_e$ rate and the low-background non-$\bar{\nu}_e$ rate obtained in Borexino is $\sim$20:1.  The same ratio in KamLAND, as quoted from Ref.~\cite{KamLAND2}, is 73:276 or 1:4, two orders of magnitude lower than in Borexino.

The results for the geo--neutrinos rate, summarized in Tab.~\ref{Tab:Results}, hint at a higher rate for geo--$\bar{\nu}_e$ than current BSE predicts.  However, the present uncertainty prevents firm conclusions.  Given the very low background achieved in Borexino, a larger exposure will yield a smaller uncertainty and more definitive conclusions: we plan to accumulate at least an exposure of 1,000\,tons$\cdot$yr, which should result in a reduction of the error by a factor of two. 

Finally, we investigate the hypothesis of a geo-reactor with a typical power of 3-10~TW at the Earth's core~\cite{Herdon}.  We assume an anti-neutrino spectrum as detailed above, with power fractions of the fuel components as from Ref.~\cite{Herdon2}:
\begin{equation}
\mbox{$^{235}$U}:\mbox{$^{238}$U} \simeq 0.75:0.25
\end{equation}
We set an upper bound for a 3\,TW geo-reactor at 95\%~C.L. by comparing the number of expected (from reactors + geo-reactor and background) and measured events in the reactor--$\bar{\nu}_e$ energy window.  Previously, this hypothesis had been studied with KamLAND data~\cite{KamLAND2,Lisi} obtaining a limit of 6.2\,TW with a 90\% C.L.

This work was funded by INFN (Italy), NSF (U.S., Grant NSF-PHY-0802646), BMBF (Germany), DFG (Germany, Grant OB160/1-1 and Cluster of Excellence ``Origin and Structure of the Universe''), MPG (Germany), Rosnauka (Russia, RFBR Grant 09-02-92430), and MNiSW (Poland).  This work was partially supported by PRIN 2007 protocol 2007 JR4STW from MIUR (Italy).  We thank Y.A.~Sokolov and J.~Mandula from IAEA for usefull discussions and consultations.  We also thank G.~Alimonti (INFN, Milano) and E.~Padovani (Politecnico di Milano) for information about MOX reactors and E.~Vrignaud for providing us with detailed information on the French EDF reactors.  We thank F.~Mantovani and F.~Vissani (INFN, LNGS) for useful discussions.

\end{document}